\documentclass[10pt,a4paper,twocolumn,nofootinbib]{revtex4}
\usepackage[section]{placeins}
\usepackage{bm}
\usepackage{graphicx,epsf}
\usepackage{pstricks}
\usepackage{color}
\usepackage{arydshln}

\def\cs2{c_{\rm{s}}^2}

\newcommand\mpl{m_{\rm Pl}}

\renewcommand\({\left(}
\renewcommand\){\right)}

\newcommand\be{\begin{equation}}
\newcommand\ee{\end{equation}}
\newcommand\bea{\begin{eqnarray}}
\newcommand\eea{\end{eqnarray}}
\newcommand\eq[1]{Eq.~(\ref{#1})}

\newcommand\fig[1]{Fig.~(\ref{#1})}

\newcommand\eps{\epsilon}

\begin{document}

\title{WMAP9 and the single field models of inflation}
\author{Laila Alabidi}
\email{laila@yukawa.kyoto-u.ac.jp}
\affiliation{Yukawa Institute for Theoretical Physics, Kyoto University, Kyoto 606-8502, Japan}

\begin{abstract}
Using the latest release from WMAP, I find that for a reasonable number of $e-$folds the tree-level potential with self coupling power $p=3$
is now excluded from the $2\sigma$ region, the axion monodromy model with the power $\alpha=2/3$ is now excluded from the $1\sigma$ confidence region for 
$N=47$ $e-$folds and for $N=61$. $\alpha=2/5$ is also excluded from the $2\sigma$ region for $N=61$. I also find that since the upper bound on 
the running has been reduced, a significant abundance of PBHs requires fractional powers of self-coupling in the Hilltop-type model.
\end{abstract}

\maketitle


Since WMAP  has released their $9^{\rm{th}}$ and final data run I thought it fitting to update the bounds on single field models of inflation. I show
that we have some interesting results as summarised in the abstract. This
paper is just a note, and at present I have no plans to submit it to a journal.  I refer the reader to Refs.~\cite{Lyth:1998xn, Liddle:2000cg,Covi:2000qx, Lyth:2009zz, Alabidi:2005qi, Alabidi:2006qa, Alabidi:2008ej,Alabidi:2010sf} for a more thorough explanation of the presented models of inflation as well as
the results from the previous runs of WMAP and COBE. Here I briefly state the potential and the corresponding results. I use the convention that $n_s$ is the spectral index, $n_s'$ is the running of the spectral
index, $r$ is the tensor to scalar ratio, $\eps$ is the slow roll parameter corresponding to the slope of the potential and $N$ is the number of $e-$folds defined as the
ratio of the scale factor at the end of inflation to its' value when scales of cosmological interested exited the horizon;
 definitions of these parameters can be found in the aforementioned references and any generic paper or book on inflationary cosmology.

The latest results from WMAP \cite{Hinshaw:2012fq,Bennett:2012fp} give the following bounds on the cosmological parameters at the $2\sigma$ confidence limit
\bea\label{obs}
0.95<n_s<0.98\,,\nonumber\\
r<0.13\,,\nonumber\\
-0.0483<n_s'<0.0062\,.\nonumber\\
\eea
where $r$ is evaluated with a zero $n_s'$ prior and $n_s'$ is evaluated with a zero $r$ prior. 
I have used the WMAP9 data combined with BAO and H0. 

I will be referring to small field models, by which I mean those which have a field excursion roughly less than the Planck scale, these correspond to the tree level potential
\be\label{tree}
V=V_0\left[1\pm\left(\frac{\phi}{\mu}\right)^p\right]\,,
\ee
where $\mu$ and $V_0$ are constants, $\mu\leq\mpl$ and $p$ can
be positive \cite{Linde:1981mu,Albrecht:1982wi,Linde:1984cd,Binetruy:1986ss,Banks:1995dp} 
or negative \cite{mutant}. The case $p=-4$ can also arise in certain
models of brane inflation \cite{DT}.
the exponential potential
\be\label{exp}
V=V_0\left[1-e^{-q\phi/\mpl}\right]\,,
\ee
where the value of the parameter $q$ depends on whether
\eq{exp} is derived from non-minimal inflation \cite{Salopek:1988qh} such as lifting a flat direction in SUSY via
a Kahler potential \cite{Stewart:1994ts} or from non-Einstein gravity \cite{Starobinsky:1980te,Lyth:1998xn}. This
potential also arises from assuming a variable Planck mass (see for example \cite{Spokoiny:1984bd, Bardeen:1987zb})
and from Higgs inflation (see for example \cite{Bezrukov:2007ep}).
and the logarithmic potential
\be\label{log}
V=V_0\left[1+\frac{g^2}{2\pi}\ln\(\frac{\phi}{Q}\)\right]\,,
\ee
$Q$ determines
the renormalisation scale and $g<1$ is the coupling
of the super-field which defines the inflaton to the super-field
which defines the flat directions.

\begin{figure}

\centering\includegraphics[width=0.9\linewidth, totalheight=2.5in]{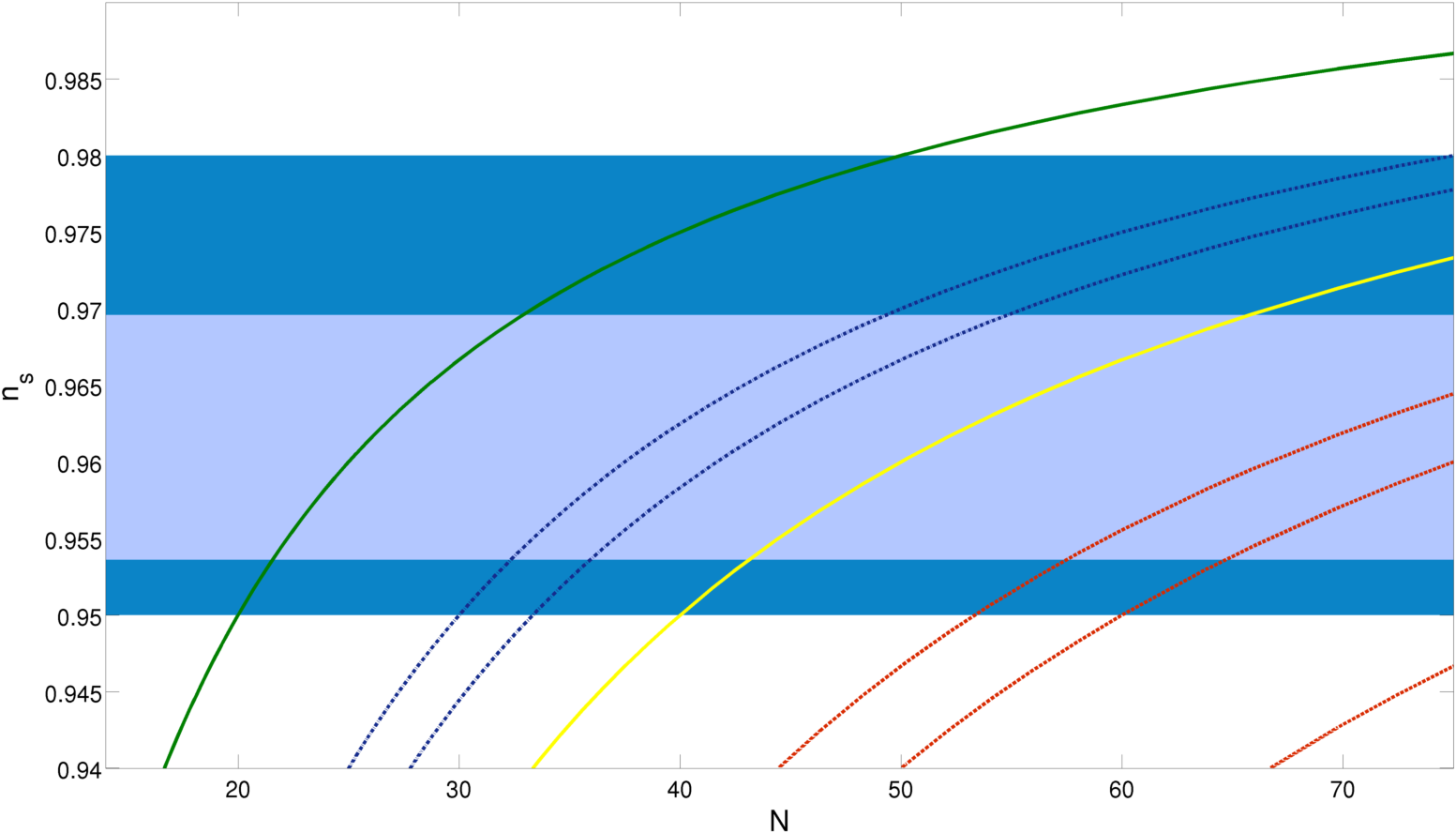}
\caption{Plot of the spectral index $n_s$ versus the number of $e-$folds $N$. The dark shaded region is the $2\sigma$ region and the light shaded region is $1\sigma$.
I have plotted the results of the small field models. The dashed red lines, bottom to top correspond to $p=3$, $p=4$ and $p=5$, the central yellow solid solid
line is $p\to\infty$, and above that are the dash-dotted dark blue lines corresponding to $p=-4$ and $p=-3$. The solid green line at the top is $p=0$.  } 
\label{fig:nvN}
\end{figure}

\begin{figure}

\centering\includegraphics[width=0.9\linewidth, totalheight=2.5in]{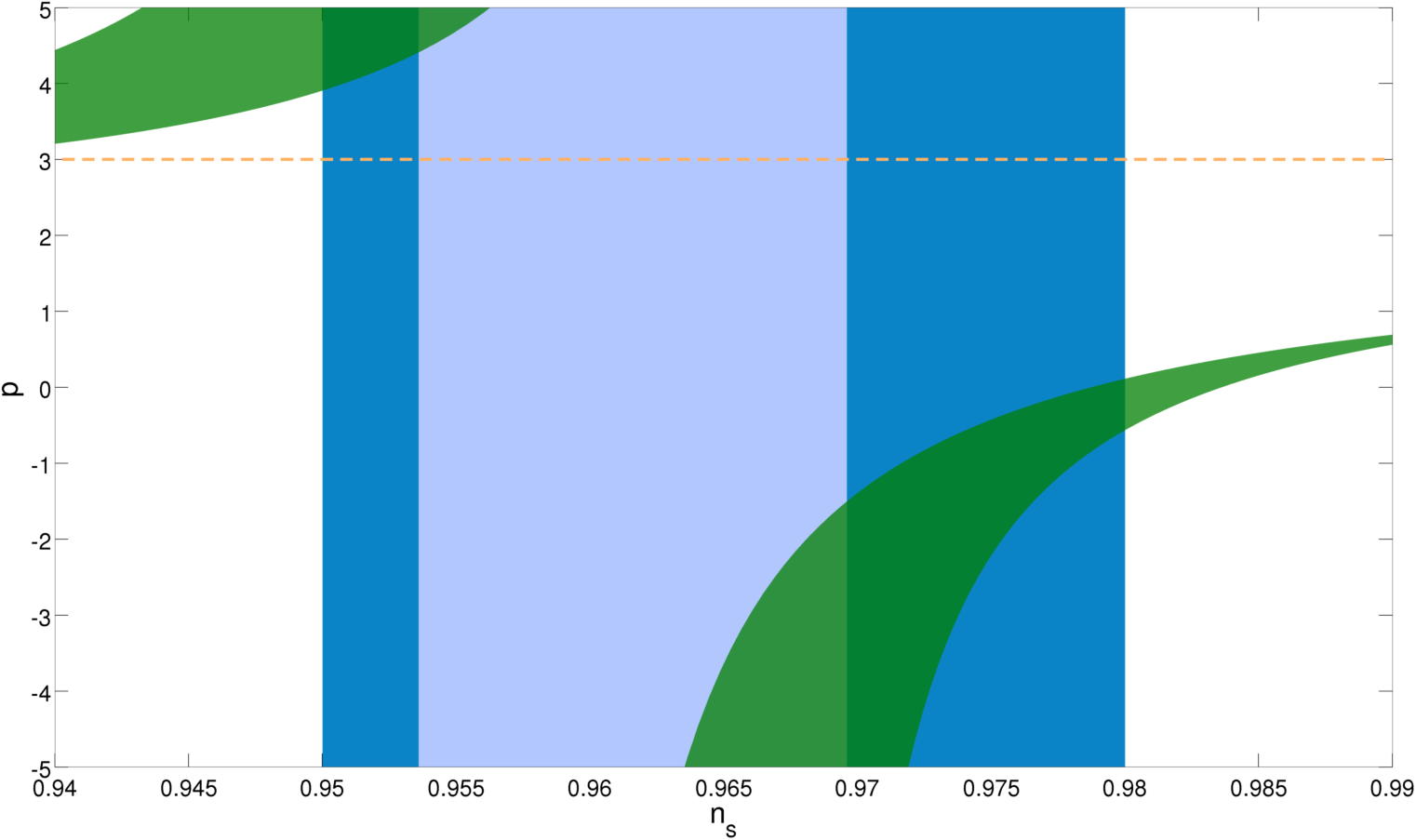}
\caption{Plot of the power of the tree level potential $p$ versus the spectral index $n_s$ for $N=54\pm7$. Dark region corresponds to $2\sigma$ while the light region is the $1\sigma$ region.
It is clear to see that $p=3$ is now ruled out.}
 \label{fig:pvn}
\end{figure}

All three models satisfy the following relation
\be\label{n_small}
1-n_s=\frac{2}{N}\(\frac{p-1}{p-2}\)\,.
\ee
with the exponential potential corresponding to $p\to\infty$ and the logarithmic potential corresponding to $p=0$. Results are plotted in \fig{fig:nvN} and \fig{fig:pvn} and it is clear to see
that $p=3$ is now excluded from the $2\sigma$ region.

The above models all predict a negligible running of the spectral index $n_s'\sim0$, however the hilltop type model \cite{Kohri:2007qn, Alabidi:2009bk} and the running mass model \cite{Stewart:1996ey,Covi:1998mb,Leach:2000ea,Covi:2002th,Covi:2004tp}
are both small field models of inflation which result in a significant positive running. This has been shown to lead to PBHs 
and a detectable spectrum of 
induced gravitational waves \cite{Saito:2009jt,Bugaev:2009zh,Alabidi:2012ex} \footnote{induced gravitational waves are gravitational waves sourced by the interaction of scalar perturbations
in the post-inflationary universe \cite{Ananda:2006af,Baumann:2007zm}}. The hilltop model has the potential:
\be\label{hill}
V=V_0\(1+\eta_p\(\frac{\phi}{\mpl}\)^p-\eta_q\(\frac{\phi}{\mpl}\)^q\)\,,
\ee
where $0<p<q$, and the Running mass model is given by
\bea\label{rmm}
V&=&V_0\left[1-\frac{\mu_0^2+A_0}{2}\(\frac{\phi}{\mpl}\)^2\right.\nonumber\\
&&\left.+\frac{A_0}{2(1+\alpha\ln(\phi/\mpl))^2}\(\frac{\phi}{\mpl}\)^2\right]\,,
\eea
where $\mu_0^2$ is the mass of the
inflaton squared, $A_0$ is the gaugino mass squared in units of $\mpl$, and $\alpha$ is related
to the gauge coupling. 

\begin{figure}

\centering\includegraphics[width=0.9\linewidth, totalheight=2.5in]{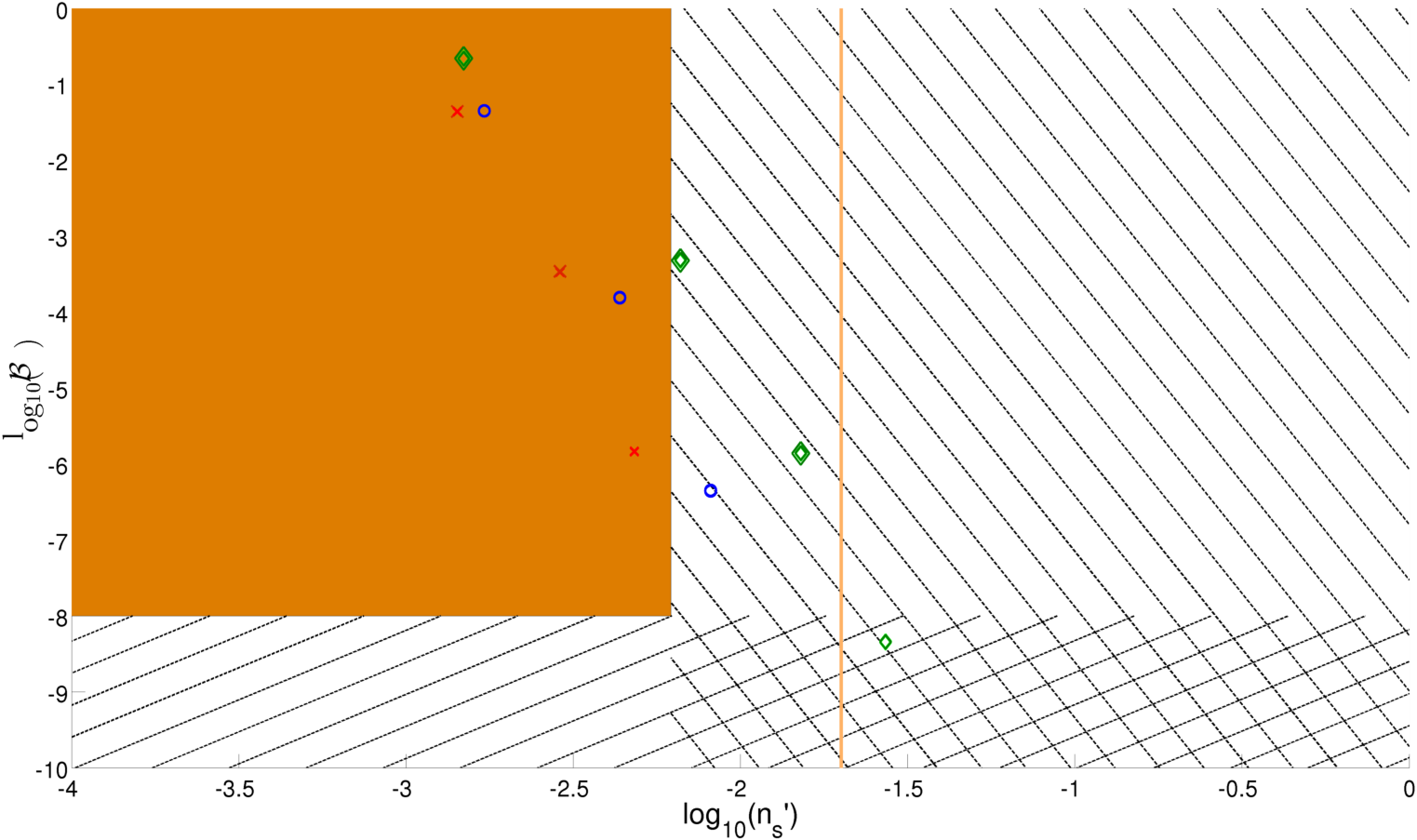}
\caption{Log-log plot of the ratio $\mathcal{B}=\eps(\phi_e)/\eps(\phi_*)$ versus the spectral running for $N=68$ $e-$folds.
The tan shaded region corresponds to $\mathcal{B}<10^{-8}$ as required by astrophysical bounds \cite{Carr:2009jm,Josan:2009qn} and $n_s'<0.0062$
as required by WMAP9. The solid pink line the upper bound on $n_s'$ from WMAP7. The red crosses $\{p,q\}=\{2,2.3\}$, the blue circles are $\{p,q\}=\{2,2.5\}$, and
the green diamonds $\{p,q\}=\{2,3\}$.}
\label{fig:hill1}
\end{figure}

I also used the parameter $\mathcal{B}\equiv\eps(\phi_e)/\eps(\phi_*)$ where the subscript $e$ refers to the end of inflation
and the subscript $*$ refers to horizon exit, and I plot the relevant results in \fig{fig:hill1}. As is clear from these figures,
the model with $\{p,q\}=\{2,3\}$ still may lead to significant Primordial Black Hole (PBH) formation, with $N=68$, but at a lower abundance than was previously thought possible. 
As we pointed in \cite{Alabidi:2010sf} changing
the bound on $n_s'$ has no impact on the running mass model as $n_s'>0.01$ coincides with inflation terminating after less than $20$ $e-$folds, 
which we dismiss as unrealistic \cite{Liddle:2003as} and the analysis for a small $n_s'$ has already been carried out.

\begin{figure}
 \centering\includegraphics[width=0.9\linewidth, totalheight=2.5in]{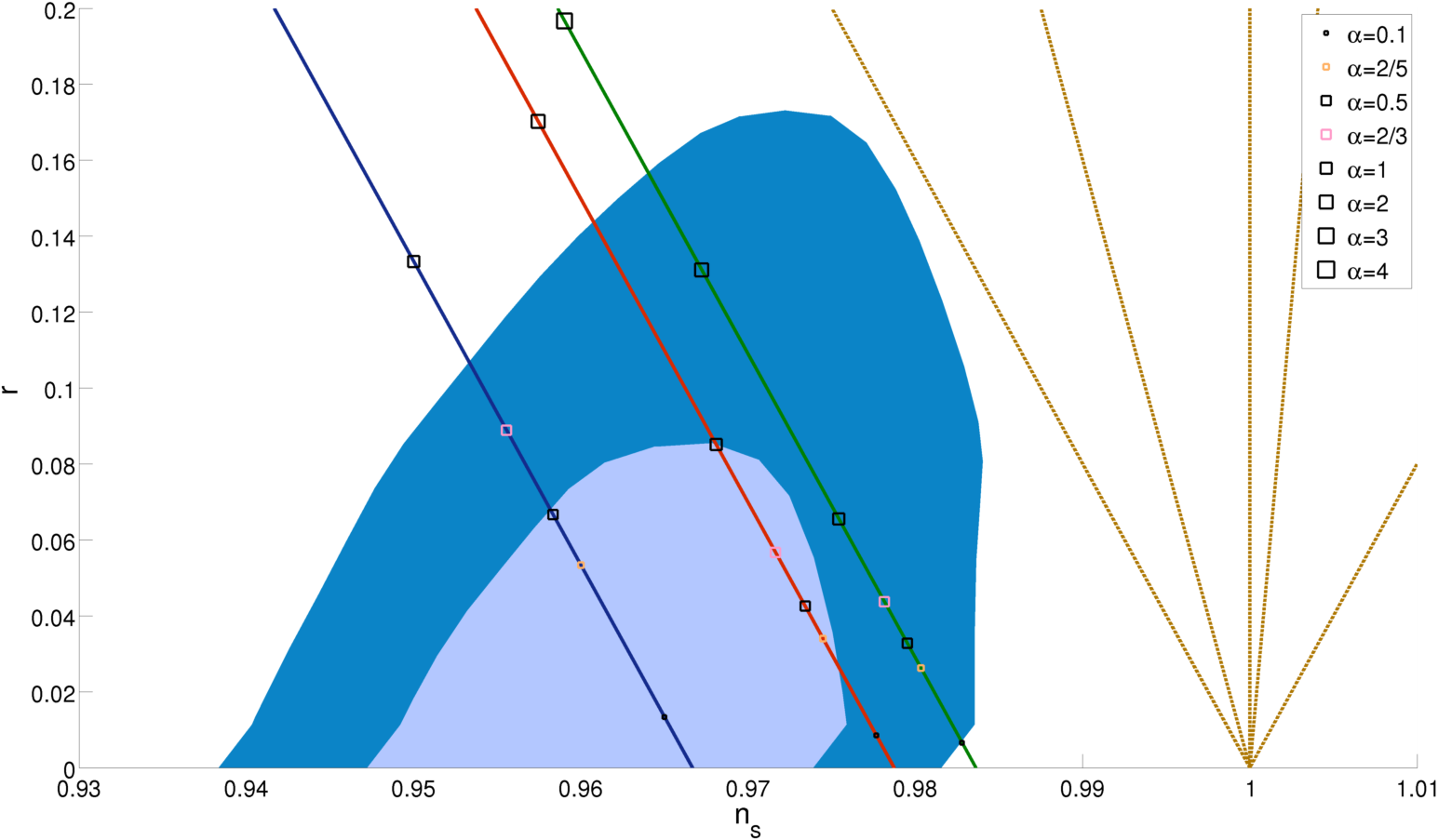}
\caption{Plot of the tensor fraction $r$ vs the spectral index $n_s$. The dark blue shaded region corresponds to the allowed region at $2\sigma$ while
the light blue region corresponds to the allowed region at $1\sigma$. The dark blue dashed lines are the WMAP 7 $2\sigma$ bound while the light blue dashed lines
are those of the WMAP7 $1\sigma$ bound. The dashed orange lines are the results for the intermediate model, with the value of $\beta$ identified in the figure. 
The solid lines are those for the chaotic potentials and $N=47$ (dark blue), $N=55$ (dark red) and $N=61$ (dark green) with the values of $\alpha$ identified in the figure.
The peach and pink boxes correspond to the axion monodromy model.}

\label{fig:big}
\end{figure}

%
Finally, the large field models have the general form:
\be
V(\phi)\propto\phi^\alpha
\ee
where $\alpha$ can be positive, as in chaotic inflation \cite{Linde:1983gd, McAllister:2008hb} or negative as in the intermediate model  \cite{Barrow:1990vx}. Results are plotted in \fig{fig:big}
and it is clear that for the intermediate model $|\alpha|=2$ is now excluded from $2\sigma$ and $|\beta\gg2$ is excluded from $1\sigma$ placing the model
under considerable strain. The axion monodromy model \cite{Silverstein:2008sg} with $\alpha=2/3$  is now excluded from the $1\sigma$ region for $N=47$ $e-$folds 
as well as for $N=61$ $e-$folds, the $\alpha=2/5$ is also excluded from the $2\sigma$ region for $N=61$.

To conclude, thanks to the WMAP9 results we can exclude the $p=3$ tree-level potential from the $2\sigma$ region, the intermediate model from the $1\sigma$ region,
and the $\alpha=2/3$ axion monodromy model from the $1\sigma$ region for $N=47$ $e-$folds. The hilltop-type model with self coupling powers $\{p,q\}=\{2,3\}$
appears not to lead to significant PBH formation within a reasonable number of $e-$folds, and we would need to consider only fractional powers of self-coupling.

Finally, with regards to single field DBI models mentioned in \cite{Alabidi:2008ej, Alabidi:2010sf}, 
we had already stated that in order to alleviate the discrepancy between theory and observation a more negative $f_{NL}^{equil}$ was needed, and so their status has not changed with the last WMAP release.

\section{Acknowledgements}
This work was supported in part by Grant-in-Aid for Scientific 
Research on Innovative Areas No.24103006.

\bibliographystyle{apsrev}
\bibliography{wmap9}
\end{document}